# Photometric Properties of Ceres from Telescopic Observations using Dawn Framing Camera Color Filters

Vishnu Reddy
Planetary Science Institute, 1700 East Fort Lowell Road, Tucson, AZ 85719, USA
Email: reddy@psi.edu

Jian-Yang Li
Planetary Science Institute, 1700 East Fort Lowell Road, Tucson, AZ 85719, USA

Bruce L. Gary
Hereford Arizona Observatory, Hereford, Arizona, USA

Juan A. Sanchez
Planetary Science Institute, 1700 East Fort Lowell Road, Tucson, AZ 85719, USA

Robert D. Stephens
Center for Solar System Studies, Landers, California, USA

Ralph Megna
Riverside Astronomical Society, Riverside, California, USA

Daniel Coley
Center for Solar System Studies, Landers, California, USA

Andreas Nathues
Max-Planck Institute for Solar System Research, Goettingen, Germany

Lucille Le Corre
Planetary Science Institute, 1700 East Fort Lowell Road, Tucson, AZ 85719, USA

Martin Hoffmann
Max-Planck Institute for Solar System Research, Goettingen, Germany

Pages: 40
Figures: 9
Tables: 5

**Proposed Running Head:** Ceres Photometry Dawn

**Editorial correspondence to:**
Vishnu Reddy
Planetary Science Institute
1700 East Fort Lowell Road, Suite 106
Tucson 85719
(808) 342-8932 (voice)
reddy@psi.edu

**Abstract**

The dwarf planet Ceres is likely differentiated similar to the terrestrial planets but with a water/ice dominated mantle and an aqueously altered crust. Detailed modeling of Ceres' phase function has never been performed to understand its surface properties. The Dawn spacecraft began orbital science operations at the dwarf planet in April 2015. We observed Ceres with flight spares of the seven Dawn Framing Camera color filters mounted on ground-based telescopes over the course of three years to model its phase function versus wavelength. Our analysis shows that the modeled geometric albedos derived from both the IAU HG model and the Hapke model are consistent with a flat and featureless spectrum of Ceres, although the values are ~10% higher than previous measurements. Our models also suggest a wavelength dependence of Ceres' phase function. The IAU G-parameter and the Hapke single-particle phase function parameter, *g*, are both consistent with decreasing (shallower) phase slope with increasing wavelength. Such a wavelength dependence of phase function is consistent with reddening of spectral slope with increasing phase angle, or phase-reddening. This phase reddening is consistent with previous spectra of Ceres obtained at various phase angles archived in the literature, and consistent with the fact that the modeled geometric albedo spectrum of Ceres is the bluest of all spectra because it represents the spectrum at 0º phase angle. Ground-based FC color filter lightcurve data are consistent with HST albedo maps confirming that Ceres' lightcurve is dominated by albedo and not shape. We detected a positive correlation between 1.1-µm absorption band depth and geometric albedo suggesting brighter areas on Ceres have absorption bands that are deeper. We did not see the "extreme" slope values measured by Perna et al. (2015), which they have attributed to "resurfacing episodes" on Ceres.

# 1. Introduction

Ceres and Vesta, the targets of NASA's Dawn mission, represent two extreme evolutionary outcomes of the planetesimal population. Vesta, an igneous body with a differentiated crust, mantle and core, experienced significant heating similar to terrestrial planets including the Earth. Ceres, on the other hand, differentiated with a much different end result where water/ice is thought to dominate its mantle and an aqueously altered curst (e.g., McCord and Sotin, 2005; Rivkin et al. 2006). Ceres contains about 1/3 of the entire mass of the asteroid belt and is 3.5 times more massive the Vesta. With just 0.3 astronomical units (AU) separating the semi major axes of these two objects in the main belt, it remains unclear how these two objects could evolve to be so dramatically different in every aspect.

Ceres has been the focus of intense Earth-based (ground-based and Hubble Space Telescope) telescopic studies since its discovery in 1801 (e.g., Ahmad, 1954; Gehrels and Owings, 1962; Tedesco et al., 1983). However, several important physical properties (surface photometric properties) remain poorly constrained. Precise understanding of photometric behavior of a surface is vital for constraining its surface properties (e.g. composition, albedo, particle size, surface roughness, etc.). Observing geometry (phase angle) affects surface albedo and spectral band parameters (band depth and slope). Overlooking these effects leads to erroneous interpretation of surface composition, space weathering, and photometric properties (Reddy et al. 2012). Neither detailed modeling of Ceres' phase function, nor the study of its wavelength dependence, has been performed so far.

Phase angle is defined as the angle between the Sun and the observer as seen from the target object. Photometric phase functions of asteroids are derived by observing the change in brightness (typically in V magnitude) as a function of phase angle from ground-based observations and have been modeled using Hapke (1981, 1984, 1986) and Lumme and Bowell (1981) scattering theories. The phase function contains important information about the physical properties of the surface, such as single-scattering albedo, particle size, packing, large-scale roughness and transparency.

An important photometric effect is the steep increase in brightness for phase angles less than ~7° (the Opposition Effect) which has been explained as being a consequence of 1) disappearing shadows at extremely low phase angle (shadow-hiding opposition effect, or SHOE), and 2) constructive interference of coherent backscattered light (coherent backscattering opposition effect, or CBOE) (e.g., Shkuratov, 1988; Hapke, 1990; Rosenbush et al., 2006; Hapke et al., 2009; Muinonen et al., 2012). The amount of the opposition effect depends on the object's albedo; low and medium albedo objects (<25%) show less prominent opposition effect and high albedo objects (E-type asteroids and Vesta) show a more obvious and narrower opposition surge (Bowell et al., 1989).

Spectral phase effects are manifested primarily as "phase reddening" and band depth changes. Phase reddening is an effect where the spectral slope of the reflectance spectrum reddens with increasing phase angle. Band depth is also affected such that increasing phase angle causes deeper absorption bands (regardless of composition) (e.g., Reddy et al. 2012). Misinterpretation of surface composition (mineral abundance and space weathering effects) is possible if one does not correct for spectral phase effects (Reddy et al. 2012).

Dawn began its survey of Ceres in April 2015. The Framing Cameras (FC) on Dawn will map the surface in seven color filters (0.4-1.0 µm) and one clear filter with a best spatial resolution of ~35-m/pixel to understand its geology and cratering history (Sierks et al. 2011). Dawn is not expected to collect any data at phase angles <7°, and most data will be at phase angles between 20° and 80° due to trajectory and orbit constraints. Being in orbit around Ceres also means that the observing geometry will have correlations with sub-spacecraft latitude to some extent depending on the orbital altitude. These effects place limitations on the accuracy of photometric modeling (Li et al., 2013). Ground-based data were planned to overcome these interpretation limitations of Dawn data. Ground based photometric data taken through the FC filters can also be used to bridge Dawn FC data with previous ground-based studies by providing the same filter set as the former and similar spatial resolution to the latter. The usefulness of this study has been demonstrated in the similar work we performed for Vesta (Reddy et al., 2012). These ground-based observations and analysis have played an important role in the Vesta phase of the Dawn mission.

## 2. Observations and Data Reduction

*2.1 Observations in 2011-13*

Photometric observations of Ceres started in 2011 and were made with the seven Dawn FC filters (Table 1), a 0.30-m Schmidt-Cassegrain telescope (SCT) at Santana Observatory (SO) (Minor Planet Center/MPC Code 646), Rancho Cucamonga, California, and a 0.11-m refractor at Goat Mountain Astronomical Research Station (GMARS) (MPC Code G79), Landers, California. Subsequent observations in 2013 were obtained with the 0.35-m SCT at SO. All our photometric data were obtained using a SBIG ST-9e CCD camera. The choice of small telescope was dictated by the fact that Ceres was too bright (~8.0 V. Mag) for larger telescopes. A total of 17,789 photometric observations of Ceres were collected in seven Dawn FC filters during two oppositions with phase angle range of 0.82°-21.4°. Observational circumstances for photometric data are shown in Table 2. In all our analyses data from both SO and GMARS is referred to as GMARS data.

Reduction and analysis of photometric data was done using Minor Planet Observer (MPO) Canopus software (Warner 2007). MPO Canopus is a Windows-based integrated software package for astrometry and photometry. Canopus is capable of reducing photometric observations of asteroids, generating lightcurves, determining their rotation period, and constructing photometric phase curves. The MPOSC3 catalog that is native to Canopus software was used for photometric analysis. The MPOSC3 catalog includes a large subset of the Carlsberg Meridian Catalog and the Sloan Digital Sky Survey. A subset of the MPOSC3 catalog consisting of stars with about the same color as the Sun and having an accuracy of ~0.05 mag for V and 0.03 mag for R was used in the reduction. Using MPOSC3 photometric accuracy of 0.02 mag is typically achieved by averaging up to 5 comparison stars per field (Warner, 2007). Differential photometry was used with night-to-night calibration of the data (generally < ±0.05 mag) using field stars converted to approximate magnitudes based on Two Micron All-Sky Survey (2MASS) J-K colors. Since no star catalog values are available for the narrow band Dawn FC filters, the nearest broadband catalog value (BVRI) was used for the comparison stars in the field.

When the brightness of Ceres is compared with uncalibrated stars in a differential photometry manner it is not possible to calculate geometric albedo; instead, only relative albedo can be derived which can at least be used in constructing a reflectivity versus rotation variation for each FC band. Since Ceres has a rotation period (9.07 hours) that is usually longer than a single night's observing session empirical adjustments were have to be made when comparing one night's light curve segment with another's in order to construct a complete phase-folded reflectivity variation.

*2.2 Observations in 2014*

Ceres was observed in 2014 in a way that was specifically designed for measuring albedo using the 7 Dawn FC filters. A 0.28-meter and a 0.35-meter Schmidt-Cassegrain telescope and SBIG ST-10XME CCD camera at the Hereford Arizona Observatory, HAO (MPC code G95) were used with the 7 FC filters to create a magnitude system for each of the filters by transferring Vega fluxes to sun-like stars located near Ceres (Table 2). These secondary standard stars could then be observed in alternation with Ceres using standard all-sky observing techniques since both were at similar elevations (air mass values) throughout each observing session.

The advantage of this all-sky photometry calibration procedure over the use of background stars for differential photometry calibration is that systematic effects related to star color sensitivity are eliminated. Differential photometry requires the use of "CCD transformation equations" to remove differences in spectral response of a telescope system using a specific filter from the spectral response standard for that filter (e.g., V-band spectral response above the atmosphere). This problem is difficult enough when using a standard filter (e.g., V-band) for obtaining magnitudes associated with that filter because of response function differences produced by the CCD's quantum efficiency (QE), telescope optical transmission and atmospheric extinction. The problem is especially difficult when the filter in use (i.e., a FC filter) differs greatly from a standard filter (e.g., a V-band filter).

By creating a magnitude system for each FC filter, using Vega as a primary standard to calibrate nearby secondary standard stars, there is no need for the use of CCD transformation equations – provided the filter width is small and atmospheric extinction differences are minimized by a proper use of all-sky observing techniques. The HAO observations adhered to all-sky requirements by alternating observations of a secondary standard star (typically 20 minutes, all filters) with Ceres observations (typically 50 minutes), following the good practices rule of beginning and ending a series of alternating observations with the standard star. This precaution minimizes the possible effect of trends in atmospheric extinction during an observing session.

The secondary standard star most used was 59 Vir, a G-type main sequence star (G0V) located a few degrees from Ceres during the 2014 observations. Because Vega was too bright for the use of reliable exposure times without saturating the CCD an aperture mask was used for Vega observations that passed ~1% of the light that was incident upon the unmasked telescope aperture. The "collecting area" ratio was measured to an accuracy of 1.8%, which is one component of uncertainty for later albedo determinations.

For each observing session data files were created using the commercially available MaxIm DL software's photometry tool that recorded the magnitude difference between the target (either Ceres or 59 Vir) and an artificial star placed in the upper-left corner of each image. These files were imported to a spreadsheet originally designed for creating light curves (for exoplanets) and modified to produce a time sequence of Ceres and 59 Vir magnitudes. An atmospheric extinction model was used to fit the 59 Vir magnitudes versus airmass, and also versus time; the model allowed for adjustment of the four major extinction components: Rayleigh scattering, ozone absorption, aerosol scattering and absorption and water vapor absorption. Total extinction was allowed to vary with time according to the need for it based on 59 Vir magnitudes. This procedure led to Ceres magnitudes corrected to "above the atmosphere."

Because a spectral energy distribution (SED) for 59 Vir was established using Vega as a primary standard, and since 59 Vir was observed in alternation with Ceres, it was possible to convert Ceres brightness measurements to fluxes for every observation.

For any star it is possible to express flux using the following general equation:

$$\text{Flux}_i \text{ [watts per m}^2 \text{ per micron]} = C_i \times 2.5119^{-\text{magnitude}}$$

Where $\text{Flux}_i$ is the flux per unit wavelength averaged over the wavelength interval of filter bandpass i, $C_i$ is a constant for this filter bandpass, and magnitude is defined zero for Vega. Since we know Vega's SED it is possible to use a magnitude difference between Vega and 59 Vir to calculate $C_i$ for i = 1 to 7 (the FC color filters bands). 59 Vir is located at RA/DE = 13:16:46.5/+09:25:27, and it is spectral type G0V with B-V = 0.644, which is similar to the sun's G2V and 0.64. Table 3 lists relevant flux and magnitude information.

The solar fluxes in Table 3 are weighted averages of the filter bandpass shapes with the "2000 ASTM Standard Extraterrestrial Spectrum Reference E-490-00." The Vega fluxes are bandpass weighted averages of the spectrum given by Glushneva et al (1992). The 59 Vir fluxes are based on our measurements on several dates of the magnitude differences between Vega and 59 Vir, which also established that 59 Vir was not variable.

The spreadsheet permitted user adjustment of a 3$^{rd}$-order model for 59 Vir magnitude versus UT for an observing session. This allowed the Ceres magnitude to be converted to Ceres flux above the atmosphere versus UT (since 59 Vir has a solar spectrum). Ceres flux was then converted to Ceres geometric albedo using the standard equation:

$$A_g [\%] = 100 \times (F_{ao} / F_s) \times [(r [au] \times d [km])/(R_a [km])]^2$$

where Fao = flux of Ceres scaled to 1 au and adjusted by a phase correction factor (asteroid phase effect using H-G function appropriate for Ceres), Fs = flux of sun at 1 au, r = distance of Ceres from the sun in au, d = distance of Ceres from Earth in km, Ra = radius of Ceres in km (470.7±3.1 km) (Thomas et al. 2005). Fao is derived from

$$F_{ao} = C_i \times 2.5119^{\text{- Ceres magnitude at opposition}}$$

Ceres magnitude at opposition = Ceres magnitude for observing date + G corr'n factor

G correction factor for phase angle ($\alpha$) = $-2.5 \times \log[(1-G) \times \varphi_1(\alpha) + G \times \varphi_2(\alpha)]$

$$\varphi_1(\alpha) = \exp\{-3.33 \times (\tan \alpha/2)^{0.63}$$

$$\varphi_2(\alpha) = \exp\{-1.87 \times (\tan \alpha/2)^{1.22}$$

The H-G function employed here is the generally accepted model by Bowell et al (1989). The phase effect correction depends on the adopted value for G, and this is one of the parameters to be solved since it should be well-constrained by observations with phase angles that range from 5.4° (near opposition) to 19.9°. Any such solution for G will require rotation phase folding using a Ceres rotation period of 9.07417 hours, with a small adjustment for changes in ecliptic longitude and prograde rotation. A more detailed description of the HAO observations and calibration procedure is given at the website: brucegary.net/Dawn/allsky.html.

## 3. Photometric Modeling

*3.1 Previous observations and models*

The phase function of Ceres was studied in great detail based on the photometric data collected through V-band from the ground for phase angles between 1.1° and 21° during its 1975-1976 opposition (Tedesco et al., 1983, T83 hereafter). An absolute *V*-band magnitude of 3.61±0.03 mag and a phase slope of 0.040±0.001 mag/deg were reported using a linear model on data at phase angles ≥7°. The U-B and B-V color at zero phase are 0.70±0.01 and 0.41±0.01, with phase coefficients of 0.0015±0.0007 and 0.0006±0.0003 mag/deg.

Helfenstein and Veverka (1989) (HV89 hereafter) applied a Hapke model fit (Hapke, 1981, 1984; 1986) to the Tedesco et al. (1983) data, assuming a roughness parameter of 20° (because the phase function at <25° phase angle is not sensitive to surface roughness). Their modeling retrieved a single-scattering albedo (SSA), $w$ = 0.057±0.004, the asymmetry factor, $g$, of the single-term Henyey-Greenstein (1pHG) single-scattering phase function = -0.40±0.01, and the amplitude, $B_0$, and width, $h$, of the shadow-hiding opposition surge of 1.6±0.1 and 0.059±0.006, respectively. The corresponding geometric albedo is 0.072. To convert the magnitude of Ceres to reflectance, they used the size of Ceres reported by Millis et al. (1987), with an equatorial radius of 479.6±2.4 km and a polar radius of 453.4±4.5 km.

Lagerkvist and Magnusson (1990) (LM90 hereafter) applied the HG phase function model (Bowell et al., 1989) adopted by IAU in 1985 (cf. ref) to ground-based data in *V*-band as part of the compilation of the Asteroid Photometric Catalog (APC) (Lagerkvist et al., 1987; 1989). The Ceres data in the APC contains data from its 1975/1976 opposition, but not the data collected by Tedesco et al. (1983). The latest update of the APC is archived in the Planetary Data System (PDS) Small Bodies Node (SBN) (Lagerkvist and Magnusson, 2011). LM90 reported an H-parameter = 3.38±0.02, and G-parameter = 0.12±0.02. Note that the H-parameter is a better representation of the absolute magnitude as the H-G model contains an empirical description of the opposition effect.

Li et al. (2006) (L06 hereafter) performed disk-integrated photometric modeling with the Hubble Space Telescope (HST), Advanced Camera for Surveys High-Resolution Channel images of Ceres acquired in its 2003/2004 opposition at phase angles of 5.4°, 6.2°, and 7.4° through three wideband filters centered at 535, 335, and 223 nm. Using the LM90 H-G model parameters, L06 derived an absolute magnitude of 3.45±0.02 mag at 535 nm, corresponding to a geometric albedo of 0.087±0.003. Note that although they used the size of Ceres with an equatorial radius of 487.3±1.8 km and a polar radius of 454.7±1.6 km derived from the same HST data (Thomas et al., 2005), the effective radius of 470.7 km (radius of a sphere with the same equatorial cross-section) is less than 1% different than the value based on the Millis et al. (1987) size. From the disk-resolved images, L06 also derived a Hapke's roughness parameter, $\theta$, of 44° (Hapke, 1984) and an SSA of 0.070±0.002 at 535 nm, although they had to assume the phase function parameters reported by HV89. The high roughness for Ceres derived by L06 is unusual.

Before showing the new data, the previous models are compared and analyzed so that we can perform reasonable comparisons with the previous photometric studies of Ceres and put our results into context.

Figure 1 shows all the previous models. It appears that the HV89 Hapke-model (red line) is about 20% lower than all other models, although it is based on the same data as the T83 model. Scaled up by 20% (not shown in the figure), the HV89 model has a much better agreement with the LM90 HG-model; scaled up by 27% (pink line), it overlaps with the T83 linear-model at phase angles >7°. We cannot find the cause for the inconsistency between the original HV89 model and other models.

Although both are based on ground-based data in the same *V*-band from multiple observatories, the T83 model and the LM90 model slightly differ in both the absolute photometric scale and the phase slope. A linear fit to the LM90 model at phase angles >7° results in a phase slope of 0.037±0.002, which is slightly shallower than the phase slope of the T83 model. On the other hand, fitting the +27%-scaled HV89 model, which is a good representation of the T83 data, with an HG-model results in an H of 3.33±0.01 and a G of 0.10±0.01, further demonstrating that the T83 data has a slightly steeper slope than the LM90 data.

Comparing the +27%-scaled HV89 model parameters (equivalent to scaling up the SSA by +27% to 0.072) and the L06 model parameters, the only significant difference is in their roughness parameters of 20° and 44°, respectively, and the difference in albedo is only about 3% (relative). The HST data appear to be consistent with both the scaled HV89 model and the L06, suggesting that the disk-integrated phase function at 5°-7° phase angles is not sensitive to phase function (Helfenstein, 1988). Calculation shows that increasing roughness from 0° to 44° while keeping other parameters fixed will decrease the total brightness of Ceres by ~0.22 mag at 25° phase angle. A further increase of roughness to 60° (the upper limit for Hapke's model assumption) causes the brightness of Ceres to decrease by another 0.2 mag.

The conclusions of this comparative analysis are: 1. Two sets of previous *V*-band photometric data of Ceres (T83 and LM90) resulted in slightly different phase functions, with their G parameters being 0.10 and 0.12 (respectively), although both resulted in similar absolute magnitudes (3.33 and 3.38), 2. The *V*-band phase function of Ceres is consistent with the following set of Hapke parameters $w$=0.070, $g$=-0.40, $B_0$=1.6, $h$=0.06, and $\theta$=20°. Disk-integrated phase function at phase angles <25° is not sensitive to roughness, which can cause an up to 0.22 mag change in the total brightness of Ceres, provided roughness is not higher than 44°.

*3.2 Comparison of our data with historical data*

Figure 2 shows our data (HAO and GMARS) collected from filter F2 (center wavelength of 555 nm) compared with historical data reported by T83 and in the APC in *V*-band. The spectrum of Ceres in the visible wavelengths is flat and featureless, enabling direct comparison of Ceres' magnitude through F2 and the standard *V*-band. Overall our data have similar phase slopes with historical data. Note that the phase slope of T83 data appears to be slightly shallower than that of the APC data, as discussed in the previous section. The overall magnitude scale of HAO data is about 0.07 mag brighter (6.7%) than the historical data at this wavelength. Also note that T83 data appear to be slightly brighter than the APC data by ~0.02 mag, which is close to the usual uncertainty of photometric data for a source of comparable brightness (7-9 mag). Since GMARS FC filter data were calibrated to the nearest UBVRI filters, similar to Reddy et al. (2012), we cannot compare their absolute photometric scale with previous data.

In almost all filters, GMARS data appear to have variable magnitude offsets from HAO data from ~+0.22 mag to ~-0.07 mag. In our modeling process that follows, we scaled GMARS data by a constant magnitude offset for each filter to achieve the best match with HAO data over the full phase angle range of the data, and to reach the best fit. This magnitude-scaling factor is equivalent to an additional free parameter in the model fitting process. The additional degree of freedom will inevitably increase the model uncertainty, but the effect is expected to be insignificant as discussed later for each case. We also performed a fit to all data (HAO + GMARS), and to HAO data only, in order to assess the impact of the uncertain broadband filter calibration of GMARS data in our modeling.

3.3 Model results

We modeled our data with the linear model using data at phase angles >7°, the HG magnitude model (Bowell et al., 1989) that was adopted by IAU, and the Hapke disk-integrated model (Hapke, 2012). The best-fit model parameters for the linear model and the HG model are listed in Table 4, and for Hapke model in Table 5. We will discuss the linear and the HG modeling in Sections 3.3.1 and 3.3.2, and the Hapke model results in Section 3.3.3. The typical model uncertainties are 0.003-0.004 for both H and G, and the total uncertainty of H is dominated by the absolute photometric calibration of our data, which is ~3%.

*3.3.1   HG model*

We used the formula given in Bowell et al. (1989) to fit an HG model for the data from each filter. The fitting is performed in a least-$\chi^2$ sense in magnitude. Figure 3 shows the model for all seven FC filters, and Fig. 4 plots our model parameters for all filters. For the H-parameter, which is an extrapolation of the observed phase function to zero phase angles and thus a measure of the absolute magnitude of the target, the results from fitting all data and HAO's data only almost exactly match each other. For the G-parameter, HAO's data dominate the best-fit values, while the values from GMARS data exhibit more scatter. One possible cause for the deviations could be due to the sporadic nature of GMARS data. As shown in Fig. 2, HAO's data usually cover a significant fraction of, or the full rotation lightcurve at each phase angle station, while GMARS data do not have such dense coverage in each rotation of Ceres. Therefore, GMARS data appear to be noisier than HAO data, resulting in larger uncertainties in the model. This is also evident from the large error bars of the G-parameters derived from GMARS data (Fig. 4). The photometric uncertainty in the GMARS data should not have significant effects on the modeled H-parameters (and geometric albedos) as suggested by the good agreement in this parameter derived using two datasets separately and using both.  In addition, in HG model, the phase slope and opposition are quantified together by a single parameter, G. Therefore the agreement in H-parameter from different datasets suggested that the scaling of the GMARS data is sufficiently good for the determination of geometric albedo.

Compared to previous HG models discussed in Section 3.1 (Table 4), our model for F2 filter has a slightly lower G-parameter (associated with a steeper phase slope) than previous models, while the H is brighter than previous models by 0.09-0.14 mag. The higher H results from both the higher brightness measurement of Ceres and the steeper phase function in our model.

The geometric albedos derived from the H-parameter, assuming an equivalent radius of 470.7 km for Ceres, are plotted in Fig. 4 and listed in Table 4. The overall trend of the geometric albedo spectrum is consistent with a flat and featureless spectrum of Ceres. The geometric albedo based on our data is ~10% higher than previous measurement (Li et al., 2006).

The G-parameter shows dependence on the wavelength, where the phase function is relatively shallower (less steep) at longer wavelengths. Similar wavelength dependence has been observed in many other objects, such as S-type asteroid (433) Eros (Clark et al., 2002) and V-type asteroid (4) Vesta (Reddy et al., 2012; Li et al., 2013). However, the physical mechanism for such a dependence on Ceres may be different from that on objects like Eros and Vesta, which show strong spectral slope in their spectra. Generally, multiply scattered light shows slower decrease in intensity with wavelength than singly scattered light, because it is more isotropic. If an object has a red spectrum, then as the albedo increases with wavelength, multiple scattering also increases, causing slower decrease of total scattered light with phase angle. Such a wavelength dependence of phase slope can explain phase reddening, a phenomenon in which the spectrum of an object reddens with increasing phase angle (Gehrels et al., 1964). However, Ceres is different with a relatively flat spectral slope, and much darker than S- and V-type asteroids. The multiple-scattering mechanism may not be the dominant cause, and a different mechanism may be at work. Note that the wavelength dependence of Ceres' phase function, G, should be associated with phase reddening, although it has never been reported before.

### 3.3.2 Linear model

The linear phase model parameters are derived by using the data at phase angles greater than 7°. The linear model parameters are listed in Table 4. Figure 3 shows the linear model for all seven FC color filter data. Similar to the HG model fit, the linear model fit is dominated by HAO data. Compared to previous models our data result in a slightly shallower phase slope (slower decrease of total brightness with phase angle), and a brighter absolute magnitude. The wavelength dependence of the linear model is similar to that in the G-parameters, where the phase slope is relatively shallower at longer wavelength. Similar to the case of HG model discussed in Section 3.3.1, the effect of the uncertain photometric scale of the GMARS data is small.

### 3.3.3 Hapke model

For the disk-integrated Hapke model, we followed the formalism as in Li et al. (2004), and adopted a 5-parameter version that includes single-scattering albedo, $w$, the asymmetry factor of a 1pHG function, $g$, the roughness parameter, $\theta$, and the amplitude and width of the shadow-hiding opposition effect, $B_0$ and $h$. We ignore coherent backscattering (Hapke, 2002) because our data at low phase angles do not allow us to constrain those parameters. The lack of data at phase angles greater than 25° does not allow us to use the 2-term Henyey-Greenstein function (e.g., McGuire and Hapke, 1995) for the single-particle phase function. We converted the magnitude measurement to radiance factor (or equivalently, I/F) using the equivalent radius of Ceres of 470.7 km, and fitted the radiance factor data in a least-$\chi^2$ sense.

In the modeling process, we investigated three different cases. First, because the effect of macroscopic roughness is small at low phase angles, and the effect of roughness on disk-integrated phase function is similar to adding backscattering into the single scattering phase function, it is not possible to constrain the roughness parameter from disk-integrated phase function within 25° phase angle (Helfenstein, 1988; Helfenstein and Veverka, 1989). We therefore have to assume a value for the roughness in our modeling. Although Li et al. (2006) reported $\theta=44°$, this value is much higher than the roughness values of almost all other asteroids and the Moon (Li et al., 2013), and the associated disk-integrated phase function model does not fit Ceres data well (Section 3.1), we considered the cases of fixing $\theta=20º$ as a value close to many other asteroids, as well as 44°. In addition, our data at small phase angles are sparse, and cannot constrain the opposition parameters well. We started with fixing the opposition parameters $B_0$=1.6 and $h$=0.06 (Helfenstein and Veverka, 1989). Therefore, the three cases are: Case 1 has $\theta=20º$, $B_0$=1.6, $h$=0.06 all fixed, and $w$ and $g$ free; Case 2 has $\theta=44º$, and all other parameters the same as in case 1; and Case 3 with $\theta=20°$ and all other parameters free to explore how well we can constrain the opposition parameters.

Similar to fitting the linear model and the HG model, the model parameters are dominated by HAO data. The $w$ and $g$ derived from different datasets are very stable, and the $B_0$ and $h$ vary and cannot be well constrained in almost all cases. The model residuals expressed in root-mean-square is 2-3% of the average measured radiance factors for all filters in all three cases, suggesting similar overall performance for all three cases in fitting the data. The uncertain photometric calibration of GMARS data should only have a relatively strong effect on $B_0$, and probably $h$. But its effects on $g$, and especially on $w$ should be smaller than the model uncertainties caused by the noise in the data. The reason is that the opposition effects and the phase function are mathematically separated in the Hapke's formulism, and are dominated by data at different ranges of phase angles. The opposition effect parameters are dominated by data at low phase angles (within the opposition surge), while the phase function is dominated by the data at moderate phase angles. We estimate the typical stochastic model uncertainties for SSA to be about 2%, for $g$ about 0.02, for $B_0$ about 0.5 and for $h$ about 0.01. The total uncertainty of SSA is dominated by the absolute photometric calibration of our data, ~3%. The combined uncertainty for the modeled geometric albedo is expected to be dominated by the model uncertainties in $w$, $B_0$, and the photometric calibration uncertainties, and are estimated to be about 10-15%. The resulting parameters derived using all data of F2 filter for the three cases are listed in Table 5, the best-fit models for F2 are plotted in Fig. 5, and all modeled parameters are plotted in Fig. 6.

Comparing the model parameters for F2 (Fig. 5), it appears that the models of all three cases agree with each other and the data for phase angles greater than 5°, whereas the Case 3 model fits the opposition effect better. The much higher roughness parameter for Cases 2 than for Case 1 results in less negative asymmetry factors at all wavelengths (Fig. 6) that are associated with less backscattering in the single-scattering phase function, demonstrating the similar effect of high roughness and more backscattering in the disk-integrated phase function at phase angles <25°. The relatively lesser backscattering for Case 2 than Case 1 results in higher SSAs to yield similar overall brightness scaling for the phase function, which is consistent with the geometric albedos. For Case 3, the best-fit opposition surge model has a higher amplitude and a narrower width than the previous HV89 model, with the SSA and asymmetry factor adjust accordingly. The much higher $B_0$ parameter and $h$ parameter at 750 nm for Case 3 are a result of a few peculiar data points near 5° phase angle, and may not be real. However, note that the opposition parameters are dominated by GMARS data, which are scaled to match the overall brightness of the well-calibrated HAO data, and the GMARS data are sparse and not corrected for Ceres' rotation. These factors could possibly introduce bias to the model. Since the model geometric albedo is sensitive to the opposition effect, the higher opposition amplitude parameter for Case 3 results in a much higher geometric albedo than Cases 1 and 2. More data at small phase angles are needed to better constrain the opposition effect and thus the geometric albedo.

Similar to the H and G parameters discussed in the previous section, the Hapke parameters also show weak wavelength dependence (Fig. 6). In all three cases (except for 750 nm in Case 3) the $g$ parameter becomes less negative with wavelength, corresponding to shallower phase functions, consistent with what the G parameters suggest. Comparing to the HV89 model, our modeled $g$ parameters at 555 nm for all three cases are either consistent, or less negative (less backscattering). The higher $B_0$ value suggests a stronger opposition effect than previously modeled. The roughness parameter cannot be well constrained from this dataset. However, a value of 44º for the macroscopic roughness is much higher than the roughness values for all asteroids and cometary nuclei previously modeled, while 20º is much more consistent with the most common situation on the surfaces of solar system small bodies (e.g., Li et al., 2013).

*3.4 Geometric albedo spectrum*

Based on our models, the geometric albedo of Ceres can be calculated. Figure 7 shows the comparison of the geometric albedos derived from various models discussed above with previous spectral and spectrophotometric measurements of Ceres. All the previous measurements, except for the value from HST, are not photometrically calibrated, and are therefore scaled to approximately match each other at 700 nm and match the HST data point at 555 nm.

The geometric albedos derived from the HG models are systematically higher than previous measurements by about 8-10%; the values from Case 1 Hapke model ($B_0$=1.6, $h$=0.06, $\theta$=20º fixed) are consistent with previous measurements; and the values from Case 3 Hapke model ($\theta$=20° fixed) are systematically higher than previous models by 10-15%. The difference between the values derived from different models is due to the difference in the opposition models, which are not well constrained by our data.

The overall shape of the geometric albedo spectrum is flat, with a slight blue slope, and bluer than all previous spectra and spectrophotometry. The difference in spectral slope might be a phase angle effect. The geometric albedo spectrum is the result of modeling to zero phase angle. The "Vilas 1" spectrum was acquired at 5.6º (Vilas and McFadden, 1992); the "Vilas 2" spectrum was acquired at 20.2° (Vilas et al., 1993); the "S3OS2" spectrum was collected at 16° phase angle (Lazzaro et al., 2004); the SMASS II spectrum was collected at 17°-20° phase angle (Bus and Binzel, 2002); and the 24-color survey was an average of several observations spanning a range of phase angles (Chapman and Gaffey, 1979). It is evident that the spectral slopes, although all generally flat, are redder for higher phase angles, while the geometric albedo spectrum has the bluest slope of all spectra. This "phase reddening" is in fact consistent with the slightly shallower phase functions at longer wavelength that we observed (Figs. 4 and 6).

*3.5 Albedo Uncertainties*

The shortest wavelength channel, at 438 nm, has the greatest calibration uncertainty. The filter response function calculated for above the atmosphere is affected by assumptions about CCD QE spectral response, telescope corrector plate transmission function and atmospheric extinction – all of which have their greatest uncertainties at the shortest wavelengths. A crucial step in the calibration process involves multiplying the filter spectral response function with these three telescope-specific spectral response functions in order to obtain the weighting function (for above the atmosphere) used to obtain a Vega flux for each FC channel; a propagation of errors analysis shows that the 438 nm channel should exhibit the greatest uncertainty. A comparison of the FC spectrum of two solar-like secondary standard stars (calibrated using the Vega-based FC magnitude scale) shows that their 438 nm brightness is greater than expected when compared with the sun's spectrum. Solar-like stars cannot be relied upon for deriving a spectral energy distribution at wavelengths shorter than ~ 450 nm due to uncertain "line-blanketing" (approximated by 1 – flux / blackbody flux) caused by metallicity differences in same spectral type stars. If the calculated Vega flux for the 428 nm channel is too high by 4 %, for example, this would cause the geometric albedo for Ceres to also be 4% too high. It is estimated that all filters with wavelengths longer than 438 nm have a calibration uncertainty of 3.3%, while the 438 nm channel is estimated to be uncertain at the 5.0% level.

## 4. Comparison of color lightcurves and HST albedo maps

Geometric albedo lightcurves of Ceres in seven Dawn FC filters shows subtle rotational spectral variations due to changing albedo and compositional variations. Figure 8 shows the geometric albedo of Ceres in each FC filter (from HAO data) as a function of the sub-Earth longitude (SEL) as defined by Thomas et al. (2005), with longitude increasing toward east. Given Ceres' fully relaxed spheroid shape (Thomas et al. 2005), the lightcurves are primarily dominated by subtle changes in albedo rather than shape or topography. All filters show a double-peak lightcurve with higher geometric albedo regions at 120° SEL and 300° SEL and lower albedo regions at 30° SEL and 210° SEL. The 120° SEL region has higher geometric albedo than the 300° SEL region and conversely the drop in geometric albedo is more pronounced at 210° SEL than at 30° SEL.

The 555 nm geometric albedo lightcurve is consistent with that of F555W filter lightcurve from Hubble Space Telescope (HST) data (Li et al. 2006) and the synthetic lightcurve constructed using HST single scattering albedo maps through the same filter. Comparing the 555 nm lightcurve with HST albedo maps through F555W filter shows remarkable consistency between the two. The drop in geometric albedo at 30° SEL is primarily associated with lower albedo terrain labeled #7 that stretches from 15° SEL to ~90° SEL. Bright region #2 is associated with the albedo high seen in the lightcurve at 120° SEL. The albedo low at 210° SEL seen in the lightcurve is due to several lower albedo spots (#8, #9, #10, #11) in the HST map. The second hump in the 555 nm geometric albedo lightcurve at 300° SEL is associated with the broad region that stretches from 290° SEL to 10° SEL. The peak of this hump is centered ~300° SEL which coincides with feature #8 in HST map. Overall, ground-based FC color filter lightcurve data are consistent with HST albedo maps confirming that Ceres' lightcurve is dominated by albedo and not shape.

## 5. Rotational Spectral Variations

*5.1 Band Depth*

Visible spectrum of Ceres is relatively featureless with a broad absorption band centered near 1.1 µm thought to be due to magnetite, which is also found in some carbonaceous chondrites (e.g., Larson et al., 1979). No measurements of rotational variations in the intensity of the absorption feature have been made so far. Band depth is a non-diagnostic spectral parameter that is affected by a range of factors including lunar-style space weathering, metal abundance, carbon abundance, particle size, phase angle and temperature (Reddy et al. 2012). Using our seven color FC spectrum we observed a systematic variation in band depth as a function of sub-Earth longitude (Fig. 9A). The data in this plot are normalized at 829 nm band and shows changes in the geometric albedo values at 917 nm. Comparing this change in band depth as a function of SEL with geometric albedo lightcurve in 749 nm band (Fig. 9B) shows a positive relationship between albedo and band depth as a function of SEL. As the geometric albedo increases (at 90° SEL) due to feature #2 (Fig. 8), the absorption band depth increases suggesting an increase in the abundance of the absorbing species due to feature #2. In contrast, the absorption band depth decreases when low albedo region associated with features 9-11 come into view suggesting a decrease in the abundance in the absorbing species.

*5.2 Spectral Slope*

Spectral slope is also a non-diagnostic parameter than could be affected by a range of factors similar to band depth (Reddy et al. 2012). Perna et al (2015) presented Ceres visible spectra measurements on different dates (December 2012 to January 2013) and suggested that their different slope values are due to longitude differences of disk-averaged reflectance. During this observing interval, they used five different solar analog stars to calibrate the Ceres spectra. The spectral ratios of those stars are flat suggesting any slope variations seen in Ceres data are not due to spurious solar analog slopes. Spectra taken 3 hours apart, using the same solar analog star for calibration had different spectral slopes (observations #2 and #3, in their Table 1). It is unclear if the difference in spectral slope is due to rotational spectral variations as Ceres rotated from 150 to 27 degrees longitude. During this 3 hours of rotation between their spectra #2 and #3, the reported spectral slopes (0.55 to 0.80 micron), changed from -0.70±0.23 to -3.18±0.62 [%/1000 Å] or a range of 4.44±0.72 [%/1000 Å] during their observational campaign. Apart from rotational slope variations, Perna et al. (2015) also found a spectral slope variation as a function of sub-Earth longitude as they report extremely different slope values for similar longitudes observed one month apart. For example, their observations #5 and #6 cover (in January 2013) the same longitude on the surface that were also observed in December 2012 (observations #3 and #2) and found spectral slopes of 0.05±0.42 [%/1000 Å] and 0.01±0.24 [%/1000 Å], instead of -3.18±0.62 [%/1000 Å] and -0.70±0.23 [%/1000 Å]. Since the Herschel Space Observatory had detected water vapor around Ceres months before and after these spectral slope measurements it was suggested that the slope differences were related to "resurfacing after outgassing episodes." We did not see "extreme" slope values similar to those measured by Perna et al. (2015).

Observations reported here are relevant to this discussion since they were made during two months using the same calibration star, and exhibit internal consistency on the order of 0.1% (band-to-band calibration SE). Figure 9C is a plot of spectral slope across a similar wavelength interval used by Perna et al. (2015) (0.55 to 0.83 micron vs. 0.50 to 0.80 micron), plotted versus longitude using the same slope scale of the Perna et al. (2015). The range of variation of slope versus wavelength that we observed is <1/10th of what is reported by Perna et al. (2015) (0.15% vs. 4%). The weaker relationship between spectral slope and SEL could be due to a couple of reasons. The first is averaging measurements made over two months that could have averaged-out short-term variations that were present during that time. The second possibility is that no outgassing episodes took place during our 2014 observations. Dawn observations may provide a better understanding short-term variations in spectral slope associated with outgassing events.

## 6. Summary

Photometric observations of Ceres using the Dawn FC filters provide additional insight into the dwarf planet's surface properties, including albedo and roughness, prior to the arrival of NASA' Dawn spacecraft. Our comprehensive ground-based study reveals the following:

- Geometric albedos derived from the H-parameter, assuming an equivalent radius of 470.7 km, are consistent with a flat and featureless spectrum of Ceres. The geometric albedo based on our data is ~10% higher than previous measurements (Li et al., 2006).

- Because Ceres has a flat spectrum and is dark (compared with S- and V-type asteroids) it is unlikely that multiple-scattering can influence albedo versus wavelength to produce the phase reddening that we observe, and which has previously not been reported.

- Slope parameter shows an obvious dependence on wavelength, where the phase function is shallower at longer wavelengths. Similar wavelength dependence has been observed in many other objects, such as S-type asteroid (433) Eros (Clark et al., 2002) and V-type asteroid (4) Vesta (Reddy et al., 2012; Li et al., 2013).

- Like the H and G parameters, the Hapke parameters also show weak wavelength dependence.

- Shape of the geometric albedo spectrum is flat, with a slight blue slope, and bluer than all previous spectra and spectrophotometry, with the exception of the "extreme" slope values measured by Perna et al. (2015), which they have attributed to "resurfacing episodes" on Ceres.

- The spectral slopes, although all generally flat, are redder for higher phase angles, while the geometric albedo spectrum has the bluest slope of all spectra. This "phase reddening" is consistent with the slightly shallower phase functions at longer wavelength that we observed.

- Ground-based FC color filter rotation lightcurve data are consistent with HST albedo maps confirming that Ceres' lightcurve is dominated by albedo and not shape.

- We detected a positive correlation between 1.1 µm absorption band depth and geometric albedo suggesting brighter areas on Ceres have absorption bands that are deeper.

**Acknowledgment**

This research work was supported by NASA Planetary Geology and Geophysics Program Grant NNX14AN35G.

## Tables

**Table 1:** Dawn Framing Camera filter names, band passes and FWHM used in our photometric study (Sierks et al., 2011). Filter band passes and SMASS spectra of Vesta and Ceres are also shown (right).

| Filter Designation | $\lambda_{eff}$ (nm) | FWHM (nm) |
|---|---|---|
| F8 | $438^{+10}/_{-30}$ | 40 |
| F2 | $555^{+15}/_{-28}$ | 43 |
| F7 | $653^{+18}/_{-24}$ | 42 |
| F3 | $749^{+22}/_{-22}$ | 44 |
| F6 | $829^{+18}/_{-18}$ | 36 |
| F4 | $917^{+24}/_{-21}$ | 45 |
| F5 | $965^{+56}/_{-29}$ | 86 |

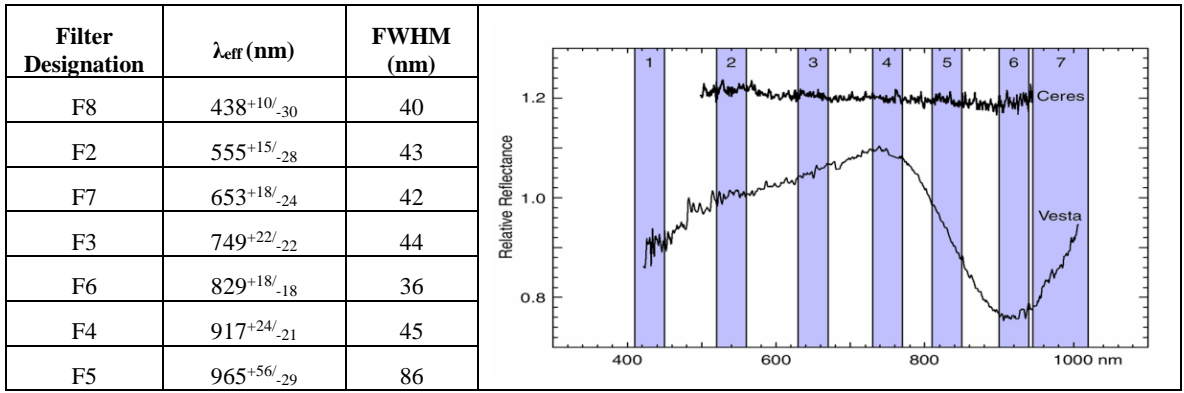

**Table 2:** Observational circumstances for Ceres observations between 2011-2014.

| Date | Hrs | Phase | Telescope | Date | Hrs | Phase | Telescope |
|---|---|---|---|---|---|---|---|
| 9/12/11 | 0:32 | 5.60 | 0.30-m* | 11/2/12 | 6:32 | 7.9 | 0.30-m* |
| 9/13/11 | 6:56 | 5.50 | 0.30-m* | 11/5/12 | 5:45 | 7.9 | 0.30-m* |
| 9/16/11 | 3:20 | 5.33 | 0.30-m* | 11/6/12 | 5:43 | 7.9 | 0.30-m* |
| 9/19/11 | 1:12 | 5.41 | 0.30-m* | 11/7/12 | 6:20 | 15.9 | 0.30-m* |
| 9/21/11 | 1:04 | 5.61 | 0.30-m* | 11/22/12 | 8:16 | 11.13 | 0.30-m* |
| 9/24/11 | 1:06 | 6.07 | 0.11-m** | 11/23/12 | 8:19 | 10.76 | 0.30-m* |
| 9/25/11 | 4:00 | 6.26 | 0.11-m** | 11/24/12 | 8:33 | 10.39 | 0.30-m* |
| 9/27/11 | 5:36 | 6.69 | 0.30-m* | 11/26/12 | 7:38 | 9.62 | 0.30-m* |
| 10/2/11 | 2:58 | 7.94 | 0.11-m** | 12/7/12 | 8:53 | 5.03 | 0.30-m* |
| 10/20/11 | 1:04 | 12.88 | 0.30-m* | 12/11/12 | 9:19 | 3.27 | 0.30-m* |
| 10/21/11 | 0:48 | 13.14 | 0.30-m* | 12/12/12 | 7:44 | 2.82 | 0.30-m* |
| 10/22/11 | 1:07 | 13.39 | 0.11-m** | 12/19/12 | 8:57 | 0.85 | 0.30-m* |
| 10/23/11 | 2:31 | 13.63 | 0.11-m** | 12/20/12 | 8:57 | 1.18 | 0.30-m* |
| 10/27/11 | 0:51 | 14.58 | 0.30-m* | 12/21/12 | 8:48 | 1.57 | 0.35-m** |
| 10/28/11 | 0:51 | 14.80 | 0.30-m* | 12/22/12 | 0:26 | 2.00 | 0.35-m** |
| 11/1/11 | 0:55 | 15.64 | 0.30-m* | 1/3/13 | 7:49 | 7.1 | 0.35-m** |
| 11/3/11 | 0:08 | 16.03 | 0.30-m* | 1/4/13 | 8:15 | 7.5 | 0.35-m** |
| 11/6/11 | 2:14 | 16.58 | 0.11-m** | 1/15/13 | 7:28 | 11.9 | 0.35-m** |
| 11/14/11 | 1:08 | 17.81 | 0.30-m* | 1/16/13 | 4:25 | 12.3 | 0.35-m** |
| 11/15/11 | 0:39 | 17.94 | 0.30-m* | 1/17/13 | 7:08 | 12.6 | 0.35-m** |
| 11/19/11 | 6:11 | 18.41 | 0.11-m** | 2/20/14 | 1:24 | 12.18 | 0.28-m*** |
| 11/22/11 | 0:45 | 18.71 | 0.30-m* | 3/23/14 | 0:36 | 11.19 | 0.28-m*** |
| 11/25/11 | 2:59 | 18.96 | 0.11-m** | 3/25/14 | 1:54 | 10.52 | 0.28-m*** |
| 11/26/11 | 4:01 | 19.0 | 0.11-m** | 4/14/14 | 2:54 | 5.36 | 0.28-m*** |
| 11/30/11 | 0:52 | 19.28 | 0.30-m* | 4/21/14 | 8:48 | 5.88 | 0.28-m*** |
| 12/7/11 | 0:52 | 19.55 | 0.30-m* | 4/25/14 | 7:42 | 6.8 | 0.28-m*** |
| 9/1/12 | 0:42 | 20.78 | 0.30-m* | 4/30/14 | 7:42 | 8.28 | 0.28-m*** |
| 9/2/12 | 0:41 | 20.85 | 0.30-m* | 5/8/14 | 4:06 | 10.88 | 0.35-m*** |
| 9/29/12 | 0:40 | 21.4 | 0.30-m* | 5/9/14 | 1:54 | 11.23 | 0.35-m*** |
| 9/30/12 | 0:37 | 21.37 | 0.30-m* | 5/13/14 | 6:18 | 12.52 | 0.35-m*** |
| 10/25/12 | 0:47 | 18.77 | 0.30-m* | 5/14/14 | 3:12 | 12.81 | 0.35-m*** |
| 10/27/12 | 0:36 | 18.41 | 0.30-m* | 6/12/14 | 1:54 | 19.77 | 0.35-m*** |
| 10/30/12 | 6:58 | 17.8 | 0.30-m* | 6/13/14 | 4:30 | 19.92 | 0.35-m*** |
| 10/31/12 | 7:02 | 17.7 | 0.30-m* | | | | |

*Santana Observatory
**Goat Mountain Research Station
*** Hereford Arizona Observatory

**Table 3:** Framing Camera filter band passes and calibration star flux information for Ceres observations. Flux measurement units are watts per meter$^2$ per micron.

| Filter Designation | Vega Flux = $C_i$ | 59 Vir Mag. | 59 Vir Flux | Solar Flux |
|---|---|---|---|---|
| F8 | 7.343e-8 | 5.772 | 3.606e-10 | 1737 |
| F2 | 3.584e-8 | 5.106 | 3.250e-10 | 1861 |
| F7 | 2.139e-8 | 4.747 | 2.699e-10 | 1587 |
| F3 | 1.383e-8 | 4.562 | 2.071e-10 | 1269 |
| F6 | 1.014e-8 | 4.419 | 1.731e-10 | 1068 |
| F4 | 0.774e-8 | 4.391 | 1.357e-10 | 894 |
| F5 | 0.689e-8 | 4.367 | 1.234e-10 | 801 |

**Table 4**: Best fit IAU HG model parameters and linear parameters. The values in parentheses are not directly from the original literature but calculated based on the reported models. The uncertainties for H, M(0) are ~0.03, dominated by the photometric calibration uncertainty, and the uncertainties for G are ~0.004, and for slope 0.006 mag/deg.

| Wavelength (nm) | H (mag) | G | Geometric Albedo | M(0) (mag) | Slope (mag/deg) |
|---|---|---|---|---|---|
| 438 | 4.05 | 0.058 | 0.10±0.01 | 4.39 | 0.039 |
| 555 | 3.24 | 0.076 | 0.099±0.003 | 3.58 | 0.038 |
| 653 | 2.87 | 0.091 | 0.099±0.003 | 3.21 | 0.037 |
| 749 | 2.64 | 0.094 | 0.096±0.003 | 2.98 | 0.037 |
| 829 | 2.49 | 0.10 | 0.097±0.003 | 2.83 | 0.036 |
| 917 | 2.44 | 0.096 | 0.093±0.003 | 2.78 | 0.037 |
| 965 | 2.43 | 0.11 | 0.093±0.003 | 2.76 | 0.036 |
| T83 | | | (0.073±0.002) | 3.61±0.03 | 0.040±0.001 |
| HV89 scaled | 3.33±0.01 | 0.10±0.01 | (0.095±0.001) | (3.61±0.03) | (0.040±0.001) |
| LM90 | 3.38±0.02 | 0.12±0.02 | (0.090±0.001) | (3.69±0.02) | (0.037±0.002) |
| L06 | | | 0.087±0.003 | | |

**Table 5**: Best fit Hapke parameters for 555 nm (F2) filter. The values in parentheses are assumed and kept fixed in the model fitting.

| Parameters | $w$ | $g$ | $B_0$ | $h$ | $\theta$ | $A_{geo}$ |
|---|---|---|---|---|---|---|
| Case 1 | 0.070 | -0.40 | (1.6) | (0.06) | (20) | 0.089 |
| Case 2 | 0.11 | -0.29 | (1.6) | (0.06) | (44) | 0.094 |
| Case 3 | 0.083 | -0.37 | 2.0 | 0.036 | (20) | 0.107 |
| HV89 | 0.057±0.004 | -0.40±0.01 | 1.6±0.1 | 0.059±0.006 | (20) | 0.072 |

**Figure Captions**

**Figure 1**. Phase curves based on previous photometric models of Ceres. References: T83: Tedesco et al. (1983); LM90: Lagerkvist and Magnusson (1990); HV89: Helfenstein and Veverka (1989); L06: Li et al. (2006).

**Figure 2**. The Dawn FC 555 nm filter data from both data sets (HAO and GMARS) compared with previous ground-based data (T83 and APC) in V-band. The HAO data are rotation-averaged where as the GMARS data are not. The uncertainties for HAO are smaller than the data points plotted.

**Figure 3**. The HG, Hapke, and linear model for data at for all seven FC filers.

**Figure 4**. The best-fit H, G parameters and the geometric albedos based on the HG-models from all filters. Stochastic model error bars are shown in the figure, and are smaller than the size of the symbols for most cases.

**Figure 5**. Hapke models for the three cases discussed in the text for F2.

**Figure 6**. The best-fit Hapke parameters and the modeled geometric albedos from all filters.

**Figure 7**. Our modeled geometric albedo spectra compared with previous visible spectra of Ceres from the ground. "Vilas 1" and "Vilas 2" are from the Vilas Asteroid Spectra dataset (Vilas et al., 1998; Vilas and McFadden 1992; Vilas et al., 1993). "S3OS2" is from the Small Solar System Objects Spectroscopic Survey dataset (Lazzaro et al., 2006; Lazzaro et al., 2004). "24-Color" is from 24-color asteroid spectrophotometry dataset (Chapman et al., 1993; Chapman and Gaffey, 1979) "HST" is from HST observations (Li et al., 2006); "SMASS II" is from the second SMASS survey (Bus and Binzel, 2002; 2003). All previous spectra and spectrophotometry are scaled to approximately match each other at 700 nm, and match HST geometric albedo at 555 nm.

**Figure 8**. Our geometric albedo light curves of Ceres through Dawn FC filters based on photometric data from HAO from 2014. Data from each night are represented by different color. The HST F555W filter albedo map from Li et al. (2006) is also shown. All figures use sub Earth longitude on the X-axis, defined such that longitude increases toward east.

**Figure 9.** (A) Geometric albedo spectrum of Ceres showing band depth variation as a function sub-Earth longitude. Data are normalized at 829 nm filter. The stochastic error for all wavelengths short ward of 829 nm is 0% and 0.01 % at 965 nm. The error bars are too small to be visible at this scale. Values for 429 nm filter are higher than previous ground-based measurements and attributed to calibration issues (B) Band depth change as a function of SEL and geometric albedo lightcurve at 749 nm wavelength showing a positive correlation between albedo and band depth.

Increasing albedo is associated with increasing absorption band depth of the 1.1-micron absorption feature. The stochastic error for all longitudes is 0.03 % and the error bars are too small to be visible at this scale. (C) Spectral slope as a function of SEL showing no significant trend. The stochastic error for all longitudes is 0.02 and the error bars are too small to be visible at this scale.

**Figures**

**Figure 1.**

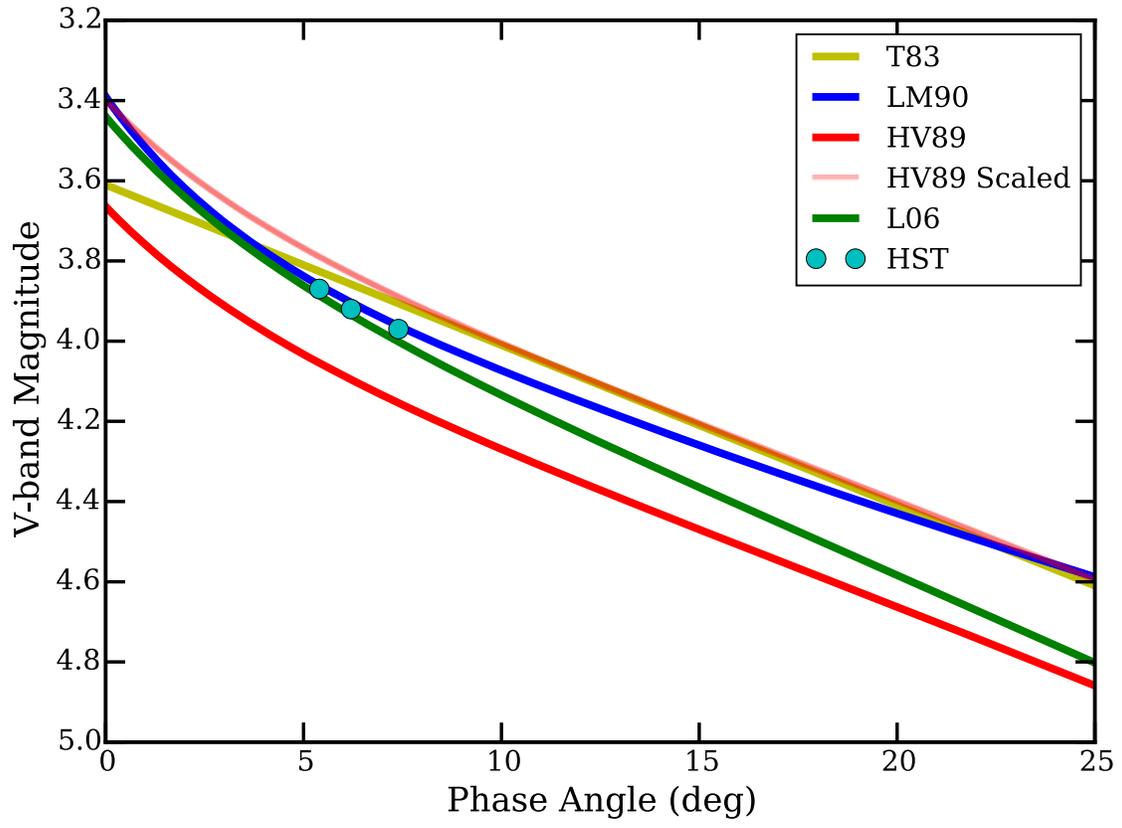

**Figure 2.**

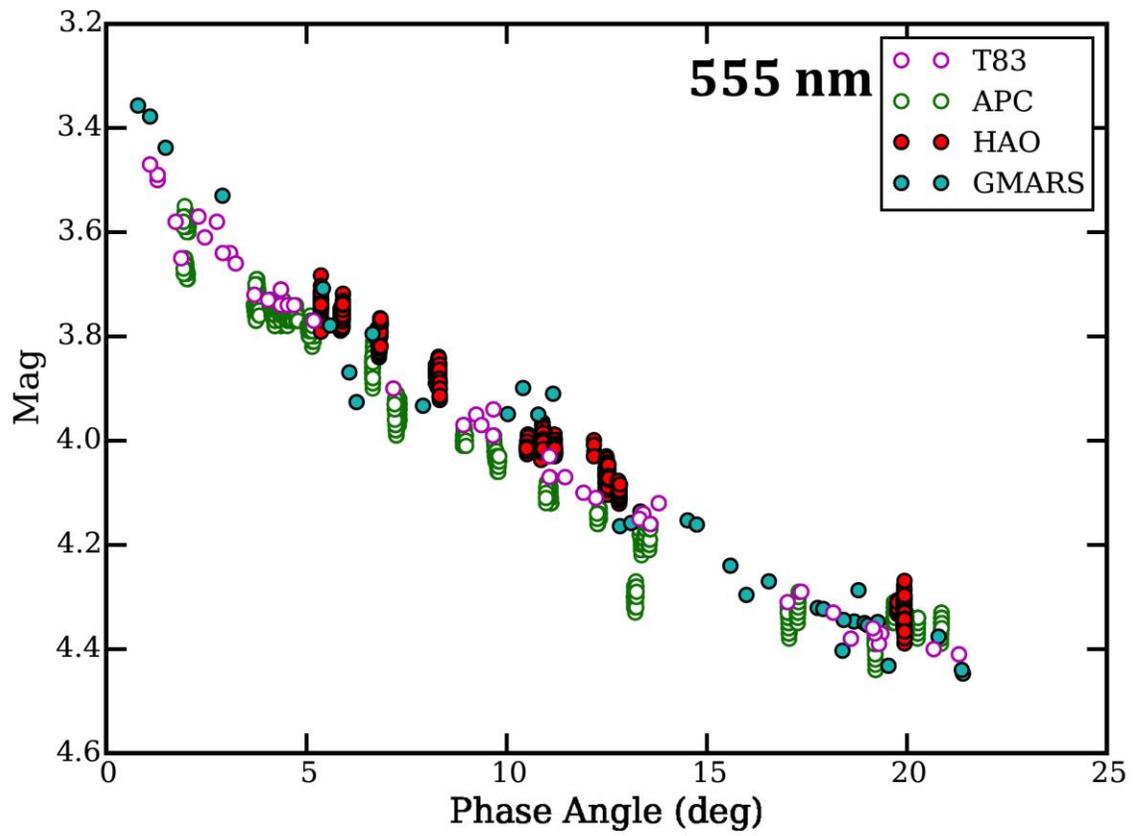

**Figure 3.**

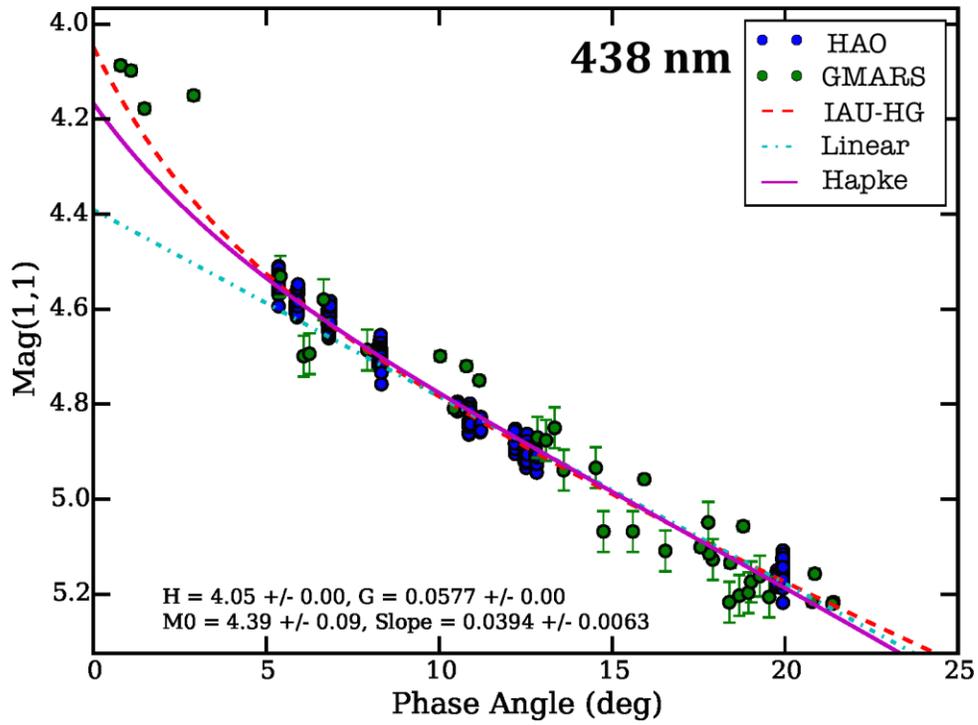

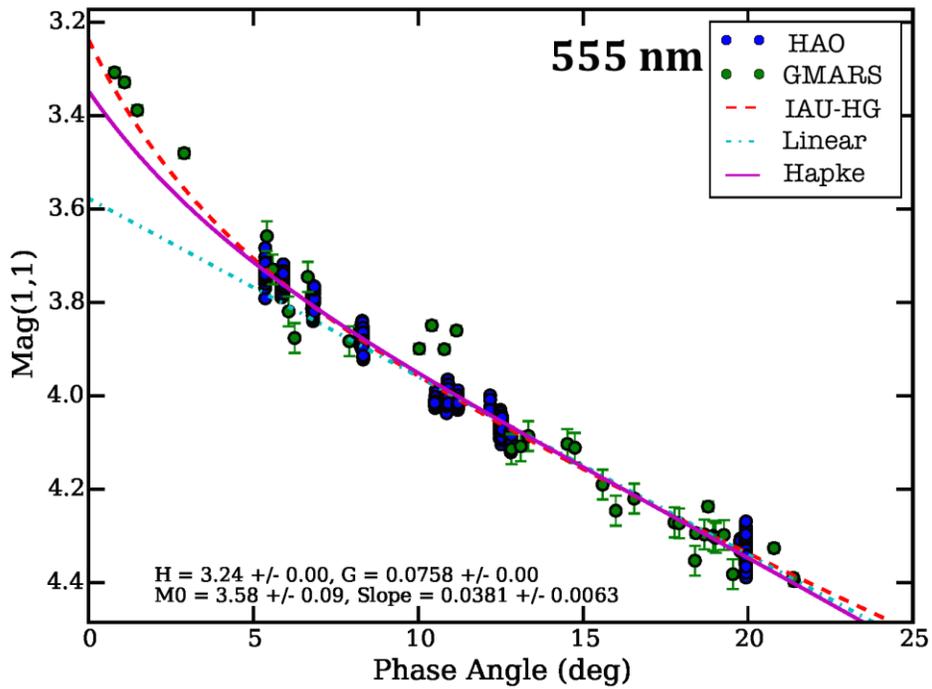

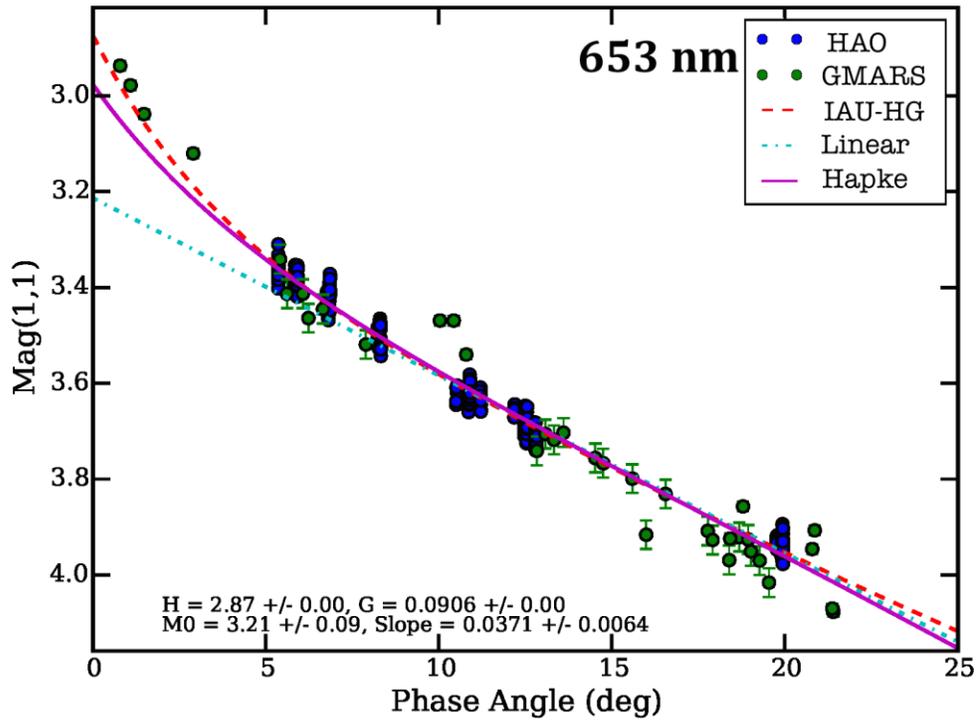

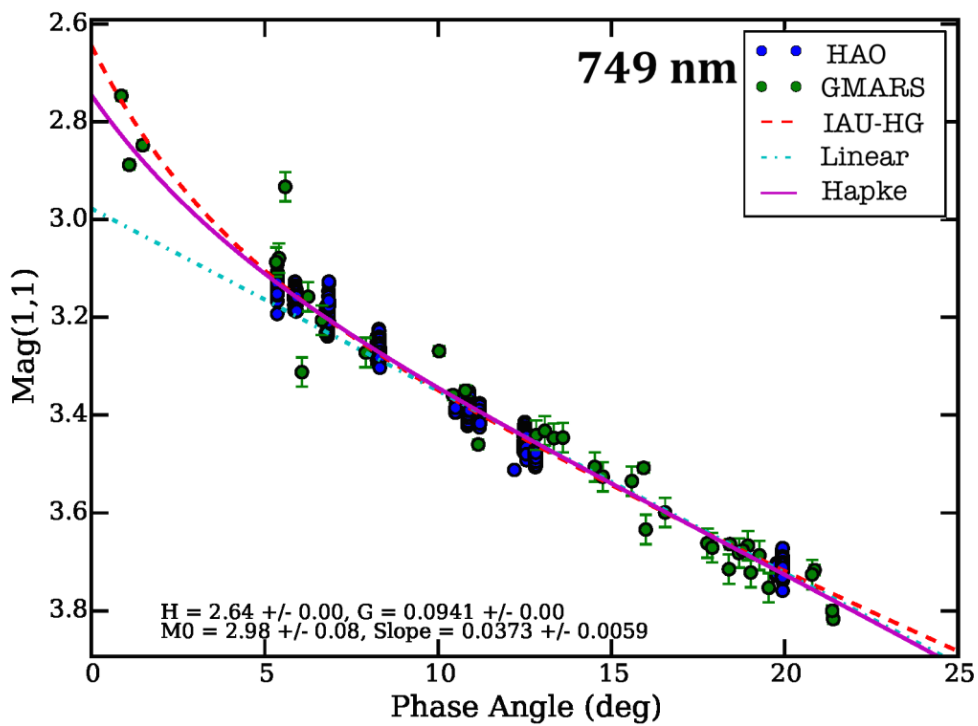

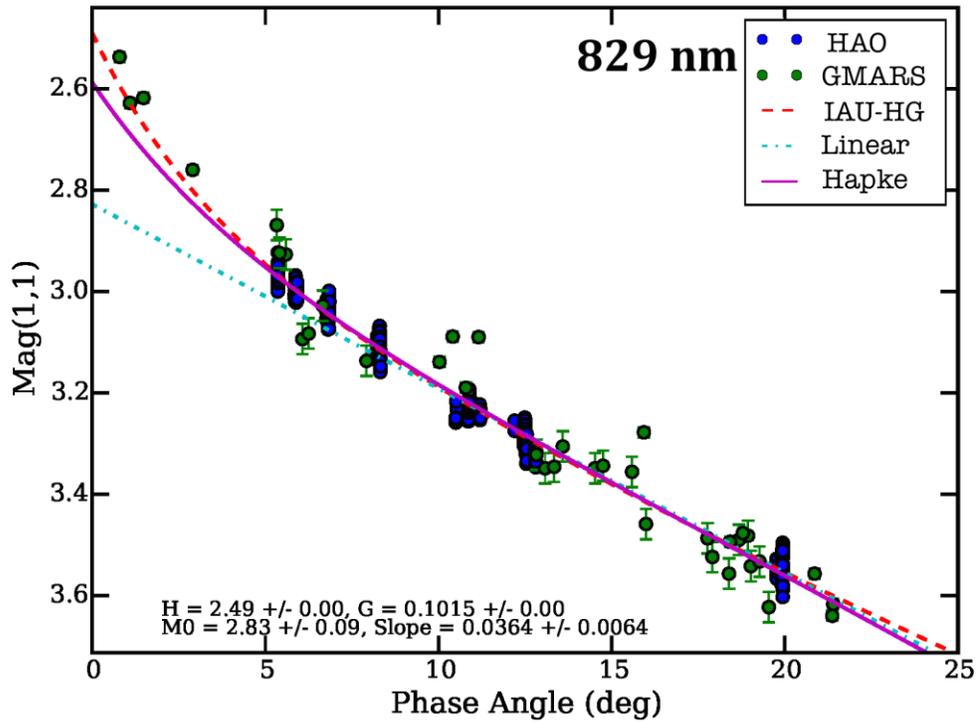
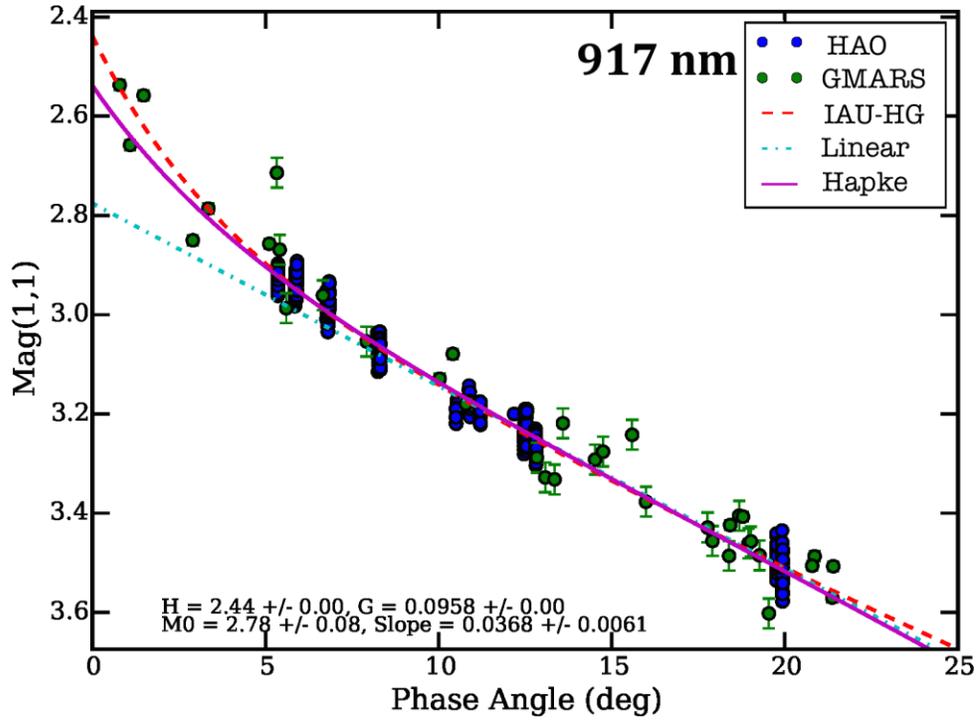

**965 nm**

H = 2.43 +/- 0.00, G = 0.1131 +/- 0.00
M0 = 2.76 +/- 0.12, Slope = 0.0359 +/- 0.0085

**Figure 4.**

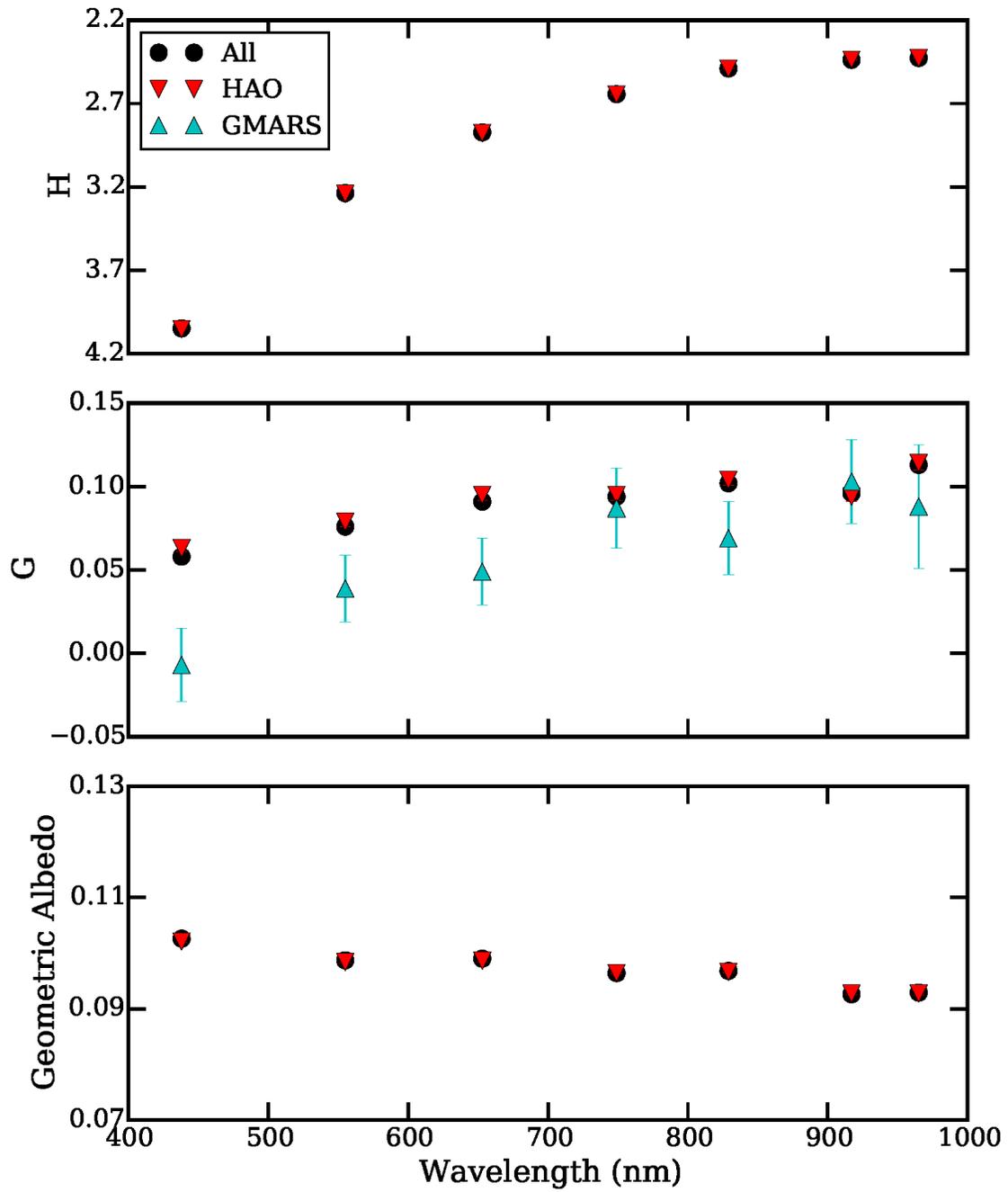

**Figure 5.**

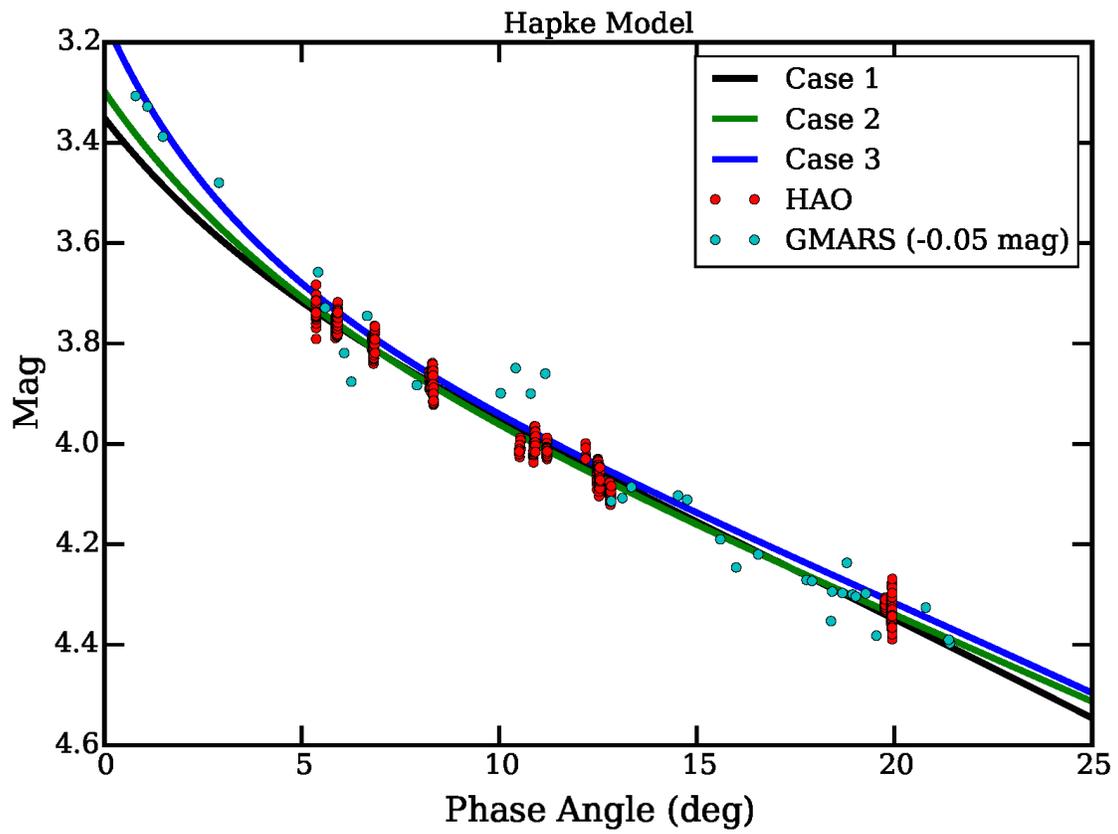

**Figure 6.**

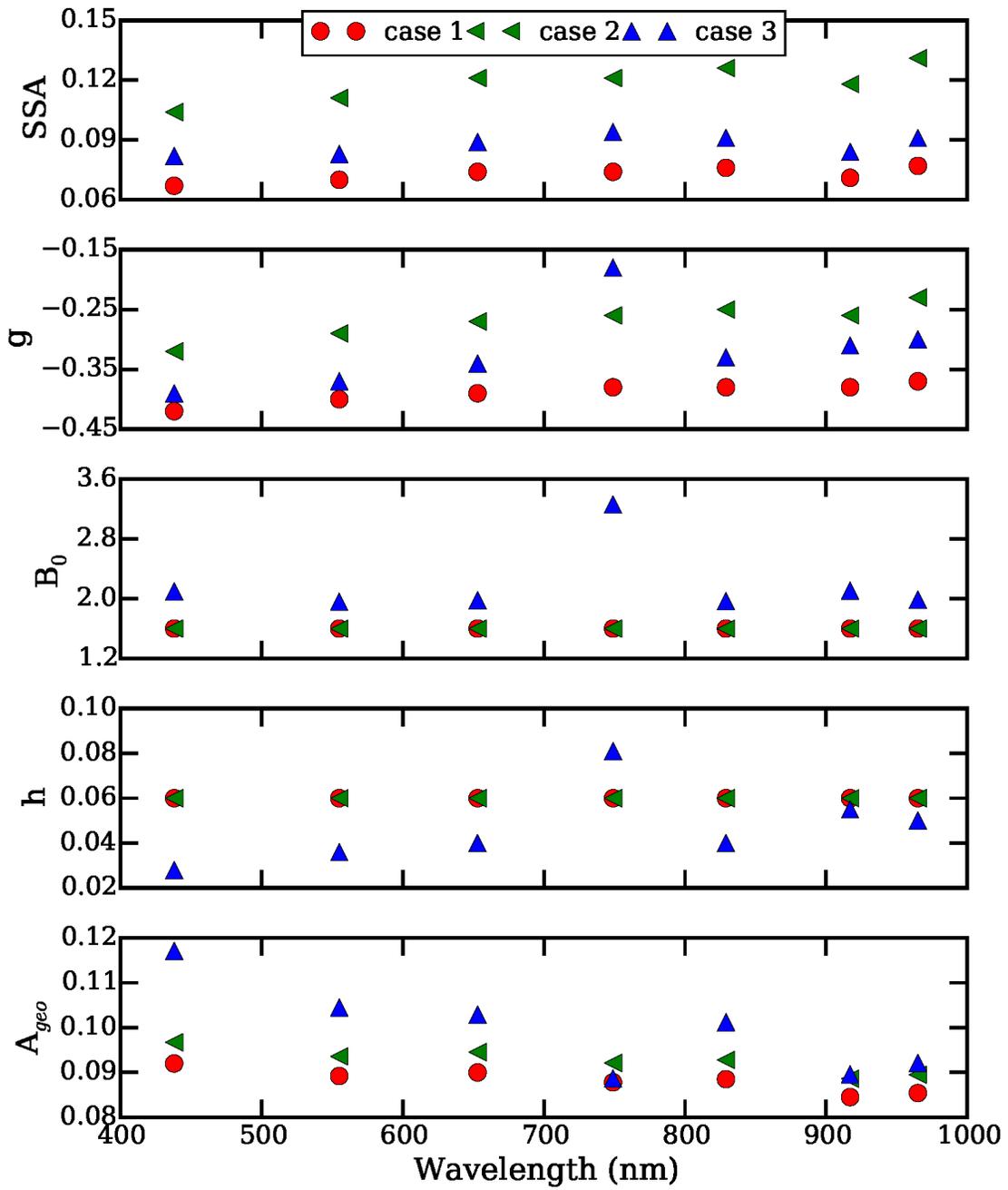

**Figure 7.**

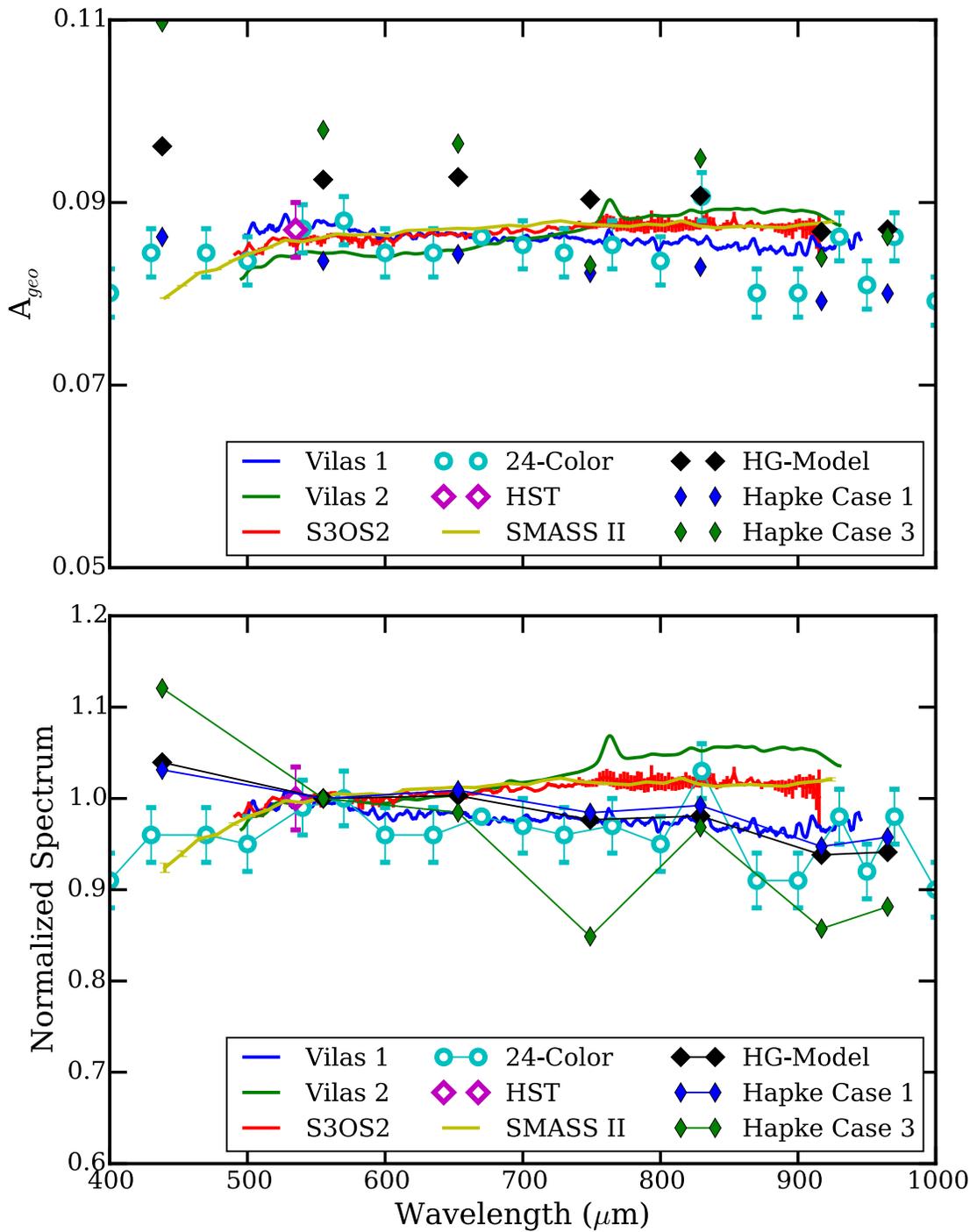

**Figure 8.**

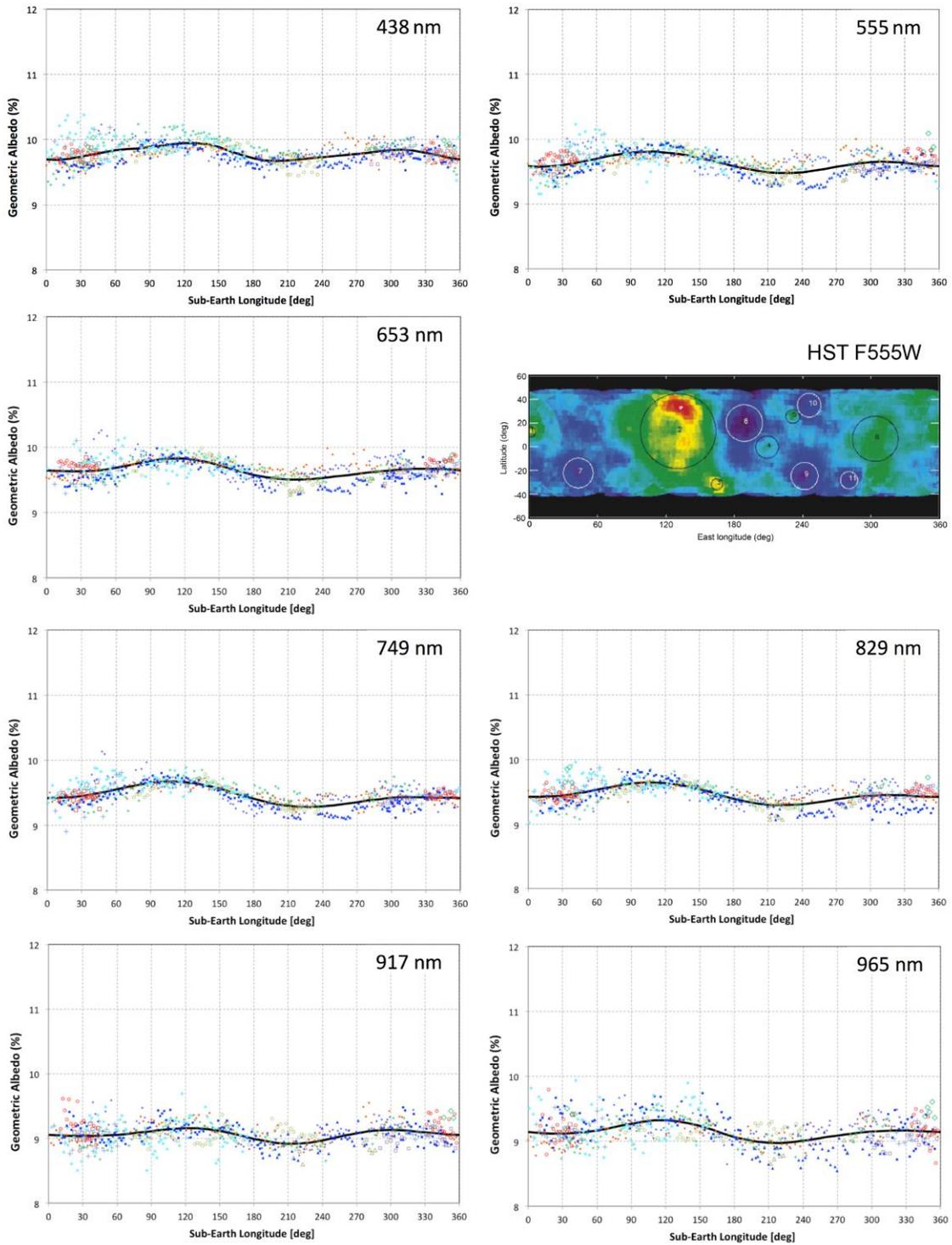

**Figure 9.**

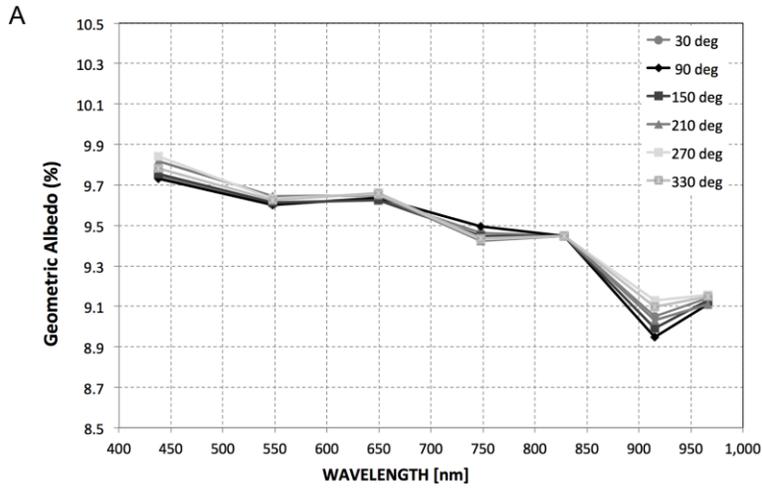

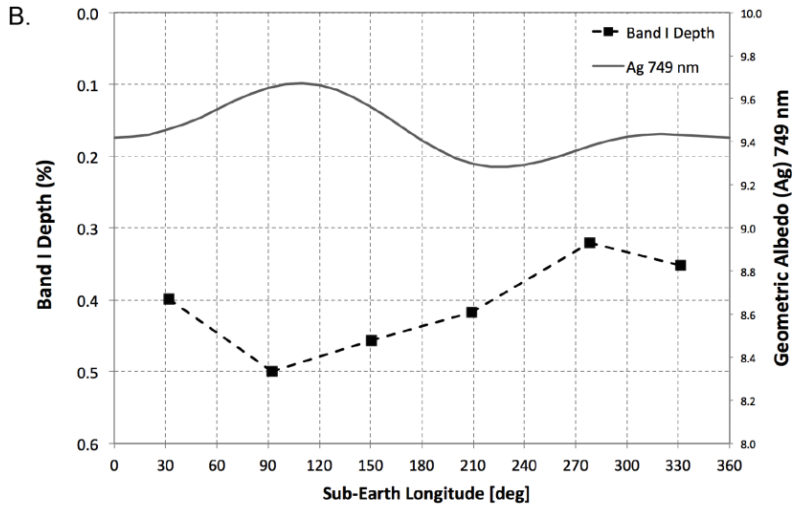

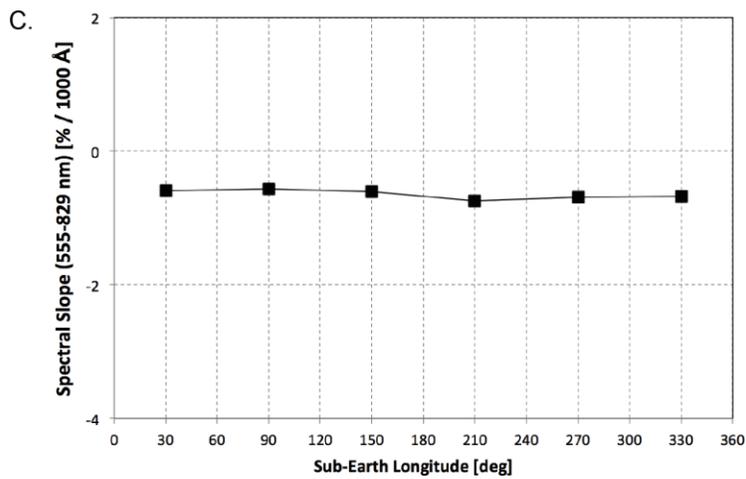